\documentclass[twocolumn]{aastex61}

\usepackage{graphicx}	
\usepackage{amsmath}	
\usepackage{amssymb}	
\usepackage{color}

\newcommand{\rockstar}{\textsc{Rockstar}}
\newcommand{\ctrees}{\textsc{Consistent Trees}}

\received{XXXX XX, XXXX}
\revised{XXXX XX, XXXX}
\accepted{XXXX XX, XXXX}
\submitjournal{ApJ}

\shorttitle{Universal scaling relation of dark matter subhalos}
\shortauthors{Hayashi et al.}

\begin{document}

\title{Universal dark halo scaling relation for the dwarf spheroidal satellites}

\correspondingauthor{Kohei Hayashi}
\email{kohei.hayashi@pku.edu.cn}

\author[0000-0002-8758-8139]{Kohei Hayashi}
\affil{Kavli Institute for Astronomy and Astrohysics, Peking University, Beijing 100871, China}
\affil{Kavli Institute for the Physics and Mathematics of the Universe (Kavli IPMU, WPI),\
 The University of Tokyo, Chiba 277-8583, Japan}

\author{Tomoaki Ishiyama}
\affiliation{Institute of Management and Information Technologies, Chiba University, 1-33, Yayoi-cho, Inage-ku, Chiba 263-8522, Japan}

\author{Go Ogiya}
\affiliation{Laboratoire Lagrange, Universit\'e C\^ote d'Azur, Observatoire de la C\^ote d'Azur, CNRS, Blvd de l'Observatoire, CS 34229, 06304 Nice cedex 4, France}

\author{Masashi Chiba}
\affiliation{Astronomical Institute, Tohoku University, Aoba-ku, Sendai 980-8578, Japan}

\author{Shigeki Inoue}
\affil{Kavli Institute for the Physics and Mathematics of the Universe (Kavli IPMU, WPI),\
 The University of Tokyo, Chiba 277-8583, Japan}
 \affil{Department of Physics, School of Science, The University of Tokyo, Bunkyo, Tokyo 113-0033, Japan}

\author{Masao Mori}
\affiliation{Center for Computational Sciences, University of Tsukuba, 1-1-1 Tennodai, Tsukuba 305-8577, Japan}



\begin{abstract}
Motivated by a recently found interesting property of the dark halo surface density within a radius, $r_{\rm max}$, giving the maximum circular velocity, $V_{\rm max}$, we investigate it for dark halos of the Milky Way's and Andromeda's dwarf satellites based on cosmological simulations.
We select and analyze the simulated subhalos associated with Milky Way-sized dark halos and find that the values of their surface densities, $\Sigma_{V_{\rm max}}$, are in good agreement with those for the observed dwarf spheroidal satellites even without employing any fitting procedures. This implies that this surface density would not be largely affected by any baryonic feedbacks and thus universal.
Moreover, all subhalos on the small scales of dwarf satellites are expected to obey the relation $\Sigma_{V_{\rm max}}\propto V_{\rm max}$, irrespective of differences in their orbital evolutions, host halo properties, and observed redshifts. Therefore, we find that the universal scaling relation for dark halos on dwarf galaxy mass scales surely exists and provides us important clues to understanding fundamental properties of dark halos.
We also investigate orbital and dynamical evolutions of subhalos to understand the origin of this universal dark halo relation and find that most of subhalos evolve generally along the $r_{\rm max}\propto V_{\rm max}$ sequence, even though these subhalos have undergone different histories of mass assembly and tidal stripping. This sequence, therefore, should be the key feature to understand the nature of the universality of $\Sigma_{V_{\rm max}}$.
\end{abstract}




\section{Introduction} \label{sec:intro}
The $\Lambda$-dominated cold dark matter ($\Lambda$CDM) theory is most successful in explaining cosmological and astrophysical observations on spatial scales larger than about 1~Mpc, including the temperature fluctuations of the cosmic microwave background~\citep[e.g.,][]{Kometal2011,Planck2015} and the clustering of galaxies~\citep[e.g.,][]{Tegetal2004}, which plays a crucial role in the formation of structure in the universe.

Nevertheless, on spatial scales smaller than 1~Mpc, i.e., galactic and sub-galactic mass scales, there are outstanding issues that the predictions from this standard theory are significantly in disagreement with some observational facts.
The representative issues include the so-called missing satellite~\citep{Klyetal1999,Mooetal1999}, core/cusp~\citep{Moore1994,Bur1995,deBetal2001,Giletal2007}, and too-big-to-fail problems~\citep{Boyetal2011,Boyetal2012}.
Whereas there are two main aspects to resolve these issues, such as baryonic physics and alternative dark matter theories, 
there still remain uncertainties in actual dark halo structure on small mass scales inferred from currently available data.

In this context, dwarf spheroidal (dSph) galaxies in the Milky Way~(MW) and Andromeda~(M31) galaxies are ideal sites for studying the nature of dark matter through internal stellar kinematics because these galaxies are the most dark matter dominated systems and are proximity that line-of-sight velocities for their resolved stars can be observed precisely by high-resolution spectroscopy~\citep[e.g.,][]{Waletal2009a,Waletal2009b}.
Moreover, these satellites have drawn special attention as the most promising targets in the indirect detection for particle dark matter though $\gamma$-ray stemmed from dark matter annihilation~\citep[e.g.,][]{Geretal2015,Hayetal2016}.
For this reason, implementing extensive dynamical analysis for those galaxies should be of great importance for shedding light on the nature of dark matter on small mass scales. 

From dynamical analysis of the galaxies including the dSphs, several studies have derived dark halo structures for various kinds of galaxies with luminosities over 14 order of magnitude based on cored dark matter density profiles with core radii ($r_0$) and densites ($\rho_0$), such as the Burkert and pseudo-isothermal profiles, and found that the central surface density, $\rho_0r_0$, is nearly constant for all these galaxies~\citep[e.g.,][]{Donetal2009,Genetal2009,Saletal2012,KF2016}.
By contrast, \citet{Boyetal2010}~defined the average column density of dark halo derived by integrating line-of-sight, and estimated it for about 300 objects ranging from dSphs to galaxy clusters. Then they concluded that the column density weakly depends on dark halo mass except for dSphs, and this trend is in agreement with the prediction of $\Lambda$CDM $N$-body simulations.

More recently, \citet[][hereafter HC15a]{HC2015a} proposed another common scale for dark halos.
They defined the surface density inside a radius of the maximum circular velocity, $\Sigma_{V_{\rm max}}$, and found that $\Sigma_{V_{\rm max}}$ shows a very weak trend or is almost constant over a wide range of $V_{\rm max}$ from dwarf galaxy to giant spiral/elliptical galaxy scales, irrespective of different dark halo profiles and types of galaxy.
In addition, \citet{HC2015b} showed that $\Sigma_{V_{\rm max}}$ is also constant with respect to $B$-band luminosity of galaxies over $\sim14$ orders of magnitude from $M_{B}=-8$ to $-22$ mag.

Following this work, \citet{OC2016}~investigated the evolution of baryonic components of the dSphs in the MW and M31, under the constraint of a constant $\Sigma_{V_{\rm max}}$ for their dark halos, and found that the models well reproduce star formation histories as derived from both observations and simulations.
Thus, the properties of the surface density $\Sigma_{V_{\rm max}}$ may play an important role in understanding star formation histories of dSph satellites and provide important constraints on the nature of dark matter and galaxy formation.

\begin{table}[t!]
    \centering
	\caption{The estimates for $V_{\rm max}$ and $\Sigma_{V_{\rm max}}$ of the eight MW and the five M31 dSphs.}
	\label{tab1}
	\footnotesize
	\begin{tabular}{lll} 
		\hline\hline
		  Object        & $V_{\rm max}$    &  $\Sigma_{V_{\rm max}}$    \\
		                & [km~s$^{-1}$]    &  [${\rm M}_{\odot}$~pc$^{-2}$]   \\
		                \hline
           {\bf Milky Way}      &&\\
            Carina         & $ 27.9^{+ 10.3 }_{- 5.7 }$   & $ 10.9^{+ 5.9 }_{- 3.7 } $ \\
			Fornax         & $ 23.3^{+ 3.8 }_{- 1.6 } $   & $ 21.0^{+ 4.6 }_{- 2.4 } $ \\
			Sculptor       & $ 24.6^{+ 3.5 }_{- 2.1 } $   & $ 28.1^{+ 9.0 }_{- 4.4 } $ \\
			Sextans        & $ 25.7^{+ 18.6 }_{- 6.9 } $  & $ 9.8^{+ 6.3 }_{- 3.2 } $ \\
			Draco          & $ 76.4^{+ 25.5 }_{- 19.6 }$  & $ 38.6^{+ 25.4 }_{- 12.1 } $ \\
			Leo~I          & $ 22.7^{+ 6.6 }_{- 3.3 } $   & $ 13.4^{+ 19.1 }_{- 6.8 } $ \\
			Leo~II         & $ 27.9^{+ 19.9 }_{- 6.6 } $  & $ 16.8^{+ 13.9 }_{- 7.9 } $ \\
			Ursa Minor     & $ 19.7^{+ 3.9 }_{- 1.9 } $   & $ 18.4^{+ 22.6 }_{- 11.1 } $ \\
           {\bf Andromeda}      &&\\
			Andromeda~I    & $ 61.3^{+ 25.8 }_{- 17.2 }$  & $ 27.7^{+ 21.9 }_{- 9.2 } $ \\
			Andromeda~II   & $ 44.3^{+ 9.0 }_{- 7.5 } $   & $ 11.9^{+ 5.3 }_{- 2.7 } $ \\
			Andromeda~III  & $ 47.7^{+ 30.8 }_{- 15.6 }$  & $ 21.3^{+ 20.9 }_{- 9.2 } $ \\
			Andromeda~V    & $ 27.3^{+ 8.8 }_{- 3.5 } $   & $ 31.1^{+ 32.6 }_{- 16.8 } $ \\
			Andromeda~VII  & $ 29.4^{+ 8.2 }_{- 3.5 } $   & $ 54.1^{+ 122.3 }_{- 33.3 } $ \\
	\hline
	\end{tabular}
\end{table}

However, our knowledge for theoretically derived $\Sigma_{V_{\rm max}}$ is yet largely uncertain due to several nonlinear effects, including the shift in inner slope of a density profile, difference in tidal effects from the host halo, and scatters in merging histories, even though the uncertainties of $\Sigma_{V_{\rm max}}$ from observations still largely remain.
In particular, calculating $\Sigma_{V_{\rm max}}$ on the dwarf galaxy mass scales in HC15a merely extrapolates from the mass-concentration relation estimated by using massive dark halos with heavier than $\sim10^{10}{\rm M}_{\odot}$ in the cosmological simulations~\citep{Klyetal2016}.
Thus, in this work, we investigate the evolution of dark matter subhalos associated with a MW-like dark halo from cosmological $N$-body simulations to understand the properties of this surface density in more details.
In particular, we focus on less massive subhalos associated with Milky Way-sized dark halos because the dark halo surface density at these halo mass scales plays a key role in understanding the nature of dark matter and formation histories of low-mass galaxies.

This paper is organized as follows. 
In Section~2, we briefly introduce a dark halo surface density defined by HC15a.
In Section~3, we describe the properties of our cosmological simulations and selection criteria of MW-sized host halos and their subhalos in this work.
In Section~4, we compare dark matter surface densities calculated from observations and simulations and then present the universal scaling relation of subhalos.
In Section~5, we discuss the origin of this universal relation by analyzing orbital evolutions of subhalos.
We summarize our results in Section~6.

\section{Dark halo surface density within a radius of maximum circular velocity} \label{sec:Sigma}
In this work, we adopt the mean surface density of a dark halo defined by HC15a to compare this density estimates from observational data with those from pure $N$-body simulations.
Given any parameters of a dark halo (e.g., scale length, scale density and any slopes of dark matter density profile), this surface density within a radius of the maximum circular velocity, $V_{\rm max}$, is given as
\begin{equation}
\Sigma_{\rm V_{\rm max}} = \frac{M(r_{\rm max})}{\pi r^2_{\rm max}},
\label{SVmax}
\end{equation}
where $r_{\rm max}$ denotes a radius of maximum circular velocity of dark halo and its enclosed mass within $r_{\rm max}$ is given as
\begin{equation}
M(r_{\rm max}) = \int^{r_{\rm max}}_{0} 4\pi\rho_{\rm dm}(r^{\prime}) r^{\prime 2} dr^{\prime}
\label{Mrmax}
\end{equation}
where $\rho_{\rm dm}$ denotes a dark matter density profile.
Under the axisymmetric assumptions, the variables of the spherical radius, $r^{\prime}$, are changed to those of the elliptical radius, $m^{\prime}$, and then Equation~(\ref{Mrmax}) can be rewritten by 
\begin{eqnarray}
M(m_{\rm max}) &=& \int^{m_{\rm max}}_{0} 4\pi\rho_{\rm dm}(m^{\prime}) Q^2m^{\prime 2} dm^{\prime},\\
m^{\prime2} &=& R^2 + \frac{z^2}{Q^2}, 
\label{Maxi}
\end{eqnarray}
where $m^{\prime}$ is described by cylindrical coordinates $(R,z)$ and dark halo axial ratio, $Q$, respectively.
This surface density is fundamentally proportional to the product of a scale density, $\rho_{\ast}$, and radius, $r_{\ast}$, for any density profiles of a dark halo, where the definition of $\rho_{\ast}$ and $r_{\ast}$ depends on an assumed density profile (in contrast to $\Sigma_{V_{\rm max}}$), such as the so-called Navarro-Frenk-White (NFW) profile~\citep{NFW1996} and Burkert profile~\citep{Bur1995}.

Calculating $\Sigma_{V_{\rm max}}$ and $V_{\rm max}$ of dark halos in the dSph galaxies, we apply our axisymmetric mass models~\citep[e.g.,][]{HC2012,HC2015b} to their available kinematic data, to obtain more reliable and realistic limits on their dark halo structures.
This is motivated by the facts that the observed light distributions of the dSphs are actually non-spherical shapes~\citep[e.g.,][]{McC2012} and $\Lambda$CDM theory predicts non-spherical virialized dark subhalos~\citep[e.g,][]{JS2002,Kuhetal2007,Veretal2014}. 
Thus, relaxing the assumption of spherical symmetry in the mass modeling should be of importance in evaluating the dark halo structures of the dSphs.

In this work, we re-estimate $\Sigma_{\rm V_{\rm max}}$ and $V_{\rm max}$ for 
the seven MW (Carina, Fornax, Sculptor, Sextans, Draco, Leo I, and Leo II) and the five M31 (And~I, And~II, And~III, And~V, and And~VII) dSphs from those estimated by HC15a and add the results of Ursa Minor dSph.
This is because we update to the latest stellar kinematic data for Draco~\citep[taken from][]{Waletal2015} and for Ursa Minor, kindly provided by Matthew~Walker (private communication), and the procedure of fitting axisymmetric mass models to their kinematic data is slightly changed from previous axisymmetric works.
Namely, in this work, we adopt an unbinned analysis for the comparison between data and models, unlike previous ones, because an unbinned analysis is the more robust method for constraining dark halo parameters rather than a binned analysis~\citep[the detailed descriptions are presented in][]{Hayetal2016}.
Estimating $\Sigma_{V_{\rm max}}$ and $V_{\rm max}$ for these dSphs, we calculate an enclosed mass within $r_{\rm max}$ along the major axis of the mass distributions.
The estimates of $\Sigma_{V_{\rm max}}$ and $V_{\rm max}$ for the total 13 dSphs are listed in Table~\ref{tab1} and plotted in Figure~\ref{fig1}.

Here, we note that the observational results from Sextans, Andromeda~II~and~VII dSphs are significantly affected by the sizes of data samples and their peculiar stellar kinematics.
Since Sextans might have a very large tidal radius, the currently available spectroscopic data for this galaxy are quite incomplete in the outer region.
Moreover, \citet{HC2015b} demonstrated that the lack of kinematic data sample in the outer region of a stellar system gives a large impact on determining dark halo parameters, especially the axial ratio of a dark halo and its velocity anisotropy (see Figure~12 in their paper).
For Andromeda~II, \citet{Hoetal2012}~first inspected the kinematical properties of this galaxy and suggested that this is a prolate rotating system, that is, their stars rotate with respect to its stellar minor axis.
This is a peculiar system having a rotation around the minor axis, to which our models cannot be properly applied.
For Andromeda~VII, although there is no observational evidence due to lack of data sample, \citet{HC2015b} suggested from the fitting results that the three dimensional stellar density distribution of this galaxy would be prefer to the prolate system.
Thus this galaxy could also have a peculiar system as well as Andromeda~II.
Therefore, there is the possibility that the obtained parameters for these dark halos may contain large systematics.

\begin{table}[t!]
    \centering
	\caption{The number of particles, the box length, the mass resolution, the softening length, and the initial redshift of the high-resolution and the Cosmogrid simulations.}
	\label{tab2}
	\footnotesize
	\begin{tabular}{cccccc} 
		\hline\hline
		                &  $N$      &     $L$        &       $m$            & $\varepsilon_{\rm DM}$ & $z_{\rm ini}$\\
		                &           & [Mpc]          & [${\rm M}_{\odot}$]   & [pc]          &\\
		                \hline
        High-res        & $2048^3$  &     $11.8$        &     $7.54\times10^3$ &   $176.5$     &   $127$ \\
        Cosmogrid       & $2048^3$  &     $30.0$        &     $1.28\times10^5$ &   $175.0$     &   $65$  \\  
	\hline
	\end{tabular}
\end{table}

\section{Dark matter simulations} \label{sec:Simulations}
\subsection{Cosmological simulation} 
In this work, we utilize the two cosmological simulations.
One is the high-resolution simulation performed by~\citet{Ishietal2016}, the another one is the Cosmogrid simulation performed by~\citet{Ishietal2013}.
The former provides the higher resolution and the latter enables us to collect more dark halo samples with the larger size of simulation box.

The run parameters in these simulations are listed in Table~\ref{tab2}, 
and the best-fit cosmological parameters are consistent with the cosmic microwave background obtained by the Planck satellite~\citep{Planck2014}, namely 
$(\Omega_0,\Omega_{\Lambda},h,n_s,\sigma_8)=(0.31,0.69,0.68,0.96,0.83)$ for the high-resolution simulation,
and~$(\Omega_0,\Omega_{\Lambda},h,n_s,\sigma_8)=(0.30,0.70,0.70,1.0,0.8)$ for the Cosmogrid simulation, respectively.

To identify halos and subhalos and their merger trees, we used \rockstar~(Robust Overdensity Calculation using K-Space Topologically Adaptive Refinement) halo finder\footnote{This finder identifies dark halos based on adaptive hierarchical refinement of friends-of-friends groups of particles in six phase-space dimensions and time.}~\citep{Behetal2013a} and~\ctrees~\citep{Behetal2013b}.
We use in this paper physical values of each dark halo computed by \rockstar~and \ctrees~analysis, namely virial mass ($M_{\rm vir}$), virial radius ($r_{\rm vir}$), the maximum circular velocity ($V_{\rm max}$), and its radius ($r_{\rm max}$).

In the high-resolution simulation, the four MW-sized dark halos are identified, where the total mass of the halos is ranging from $\sim 0.8$ to $\sim 3.0\times10^{12}{\rm M}_{\odot}$ as estimated from dynamical analysis of blue horizontal branch stars or/and dwarf satellites~\citep[e.g.,][]{Saketal2003,Deaetal2012,Kafetal2014}.
On the other hand, in order to get a number of dark halo samples, we identify the 56 MW-sized halos with a wider mass range from $\sim1.0$ to $\sim6.0\times10^{12}{\rm M}_{\odot}$ and select the 18 MW-sized halos that possess over 10 subhalos which satisfied some criteria~(as described the next section) among the 56 identified dark halos~(as seen in Figure~\ref{fig2}).

In Table~\ref{tab3}, we summarize the virial mass, virial radius, concentration indicator (as described below) and the number of subhalos satisfied the criteria (as described below) for the four MW-sized dark halos in each simulation, respectively.
For the Cosmogrid simulation, we show the four host halos showing relatively rapid growth at a high redshift out of 18 MW-like dark halos, that is, the subhalos associated with these parent halos did fall in at some earlier time and thus might have passed through the pericenter of the host several times.
Using these subhalos, we will discuss in details their orbital evolutionary histories in the Section~\ref{sec:Evolution}.

\begin{table}[t!]
	\centering
	\caption{The virial mass, the virial radius, the concentration indicator, and the number of subhalos within four MW-sized dark halos in each simulation.}
	\label{tab3}
	\footnotesize
	\begin{tabular}{ccccc} 
		\hline\hline
		Name & $M_{\rm vir}$              & $r_{\rm vir}$    & $\rho_{V_{\rm max}}$ & $N_{\rm subhalos}$\\
		     & [$\times10^{12}{\rm M}_{\odot}$] & [kpc]            & [$\times10^{4}$]     &\\		
		     \hline
		High-resolution &&&&\\     
		  H1 & $2.88$ & $374$ & $0.36$ & $187$\\
		  H2 & $2.40$ & $352$ & $0.35$ & $146$\\
		  H3 & $2.31$ & $348$ & $1.49$ & $118$\\
		  H4 & $1.11$ & $272$ & $0.61$ & $48$\\
		Cosmogrid &&&&\\     
		 CG1 & $4.28$ & $420$ & $0.13$ & $41$\\
		 CG2 & $2.87$ & $368$ & $0.44$ & $19$\\
		 CG3 & $3.36$ & $388$ & $0.77$ & $20$\\
		 CG4 & $3.08$ & $377$ & $0.75$ & $18$\\
	\hline
	\end{tabular}
\end{table}

In what follows, we investigate the properties of dark halo surface densities of subhalos within the selected host halos in each simualtion.

\begin{figure*}[t!]
	\begin{center}
	\includegraphics[scale=0.78]{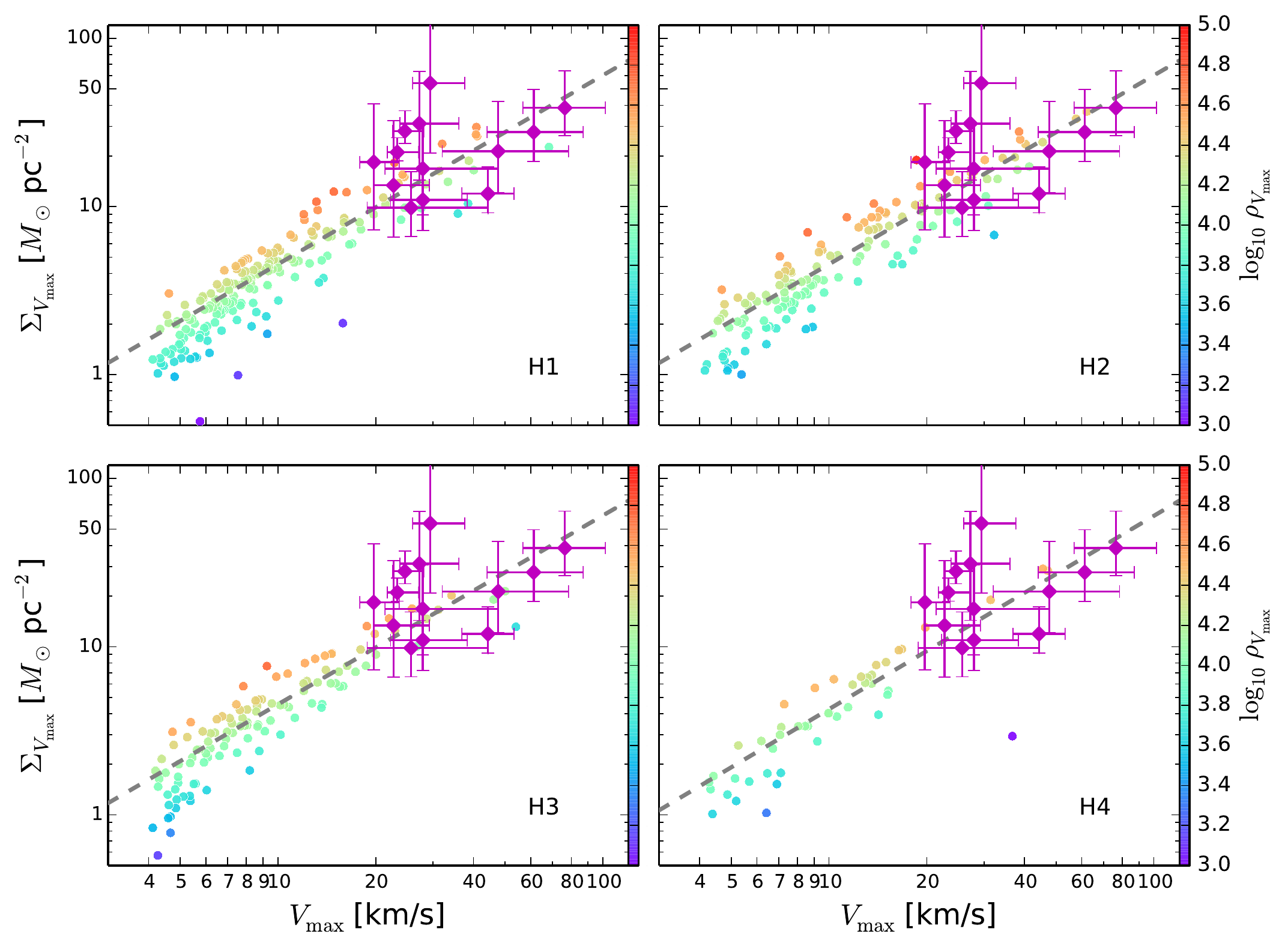}
	\end{center}
    \caption{$\Sigma_{V_{\rm max}}$ as a function of $V_{\rm max}$ focusing on dark subhalos have $V_{\rm max}\leq120$~km~s$^{-1}$. The colored dots depict the simulated subhalos associated with each MW-sized halo. Those color difference indicates $\rho_{V_{\rm max}}$. The magenta diamonds with error bars indicate the MW and M31 luminous dwarf spheroidals. The dashed lines are the fitting results of $\Sigma_{V_{\rm max}}\propto V_{\rm max}^{\alpha}$ relation (see text for more details).}
    \label{fig1}  
\end{figure*}

\subsection{Subhalo criteria} 
We extract the catalog data of subhalos within these hosts and impose criteria to avoid several numerical influences on small mass dark halos. 

Firstly, in order to avoid the effects caused by the limited spatial resolution of the simulation, we select the halos having a scale radius ($r_{\rm s}$) of a dark matter profile larger than twice the softening length of this simulation\footnote{We also perform the same calculations of dark halo properties such as $\Sigma_{V_{\rm max}}$ in which case $r_{\rm s}$ is larger than thrice the softening length, and confirm that our conclusions in this work do not largely influenced by this criterion.}.
Secondly, the virial mass is larger than $\sim10^7 {\rm M}_{\odot}$ ($\sim10^8 {\rm M}_{\odot}$) in the high-resolution~(the Cosmogrid) simulations so that a halo has at least around 1,000 dark matter particles.
Finally, we select a subhalo that settles within the radius of $r_{\rm vir}$ of a host halo at the redshift $z=0$.

\begin{figure*}[t!]
	\begin{center}
	\includegraphics[scale=0.18]{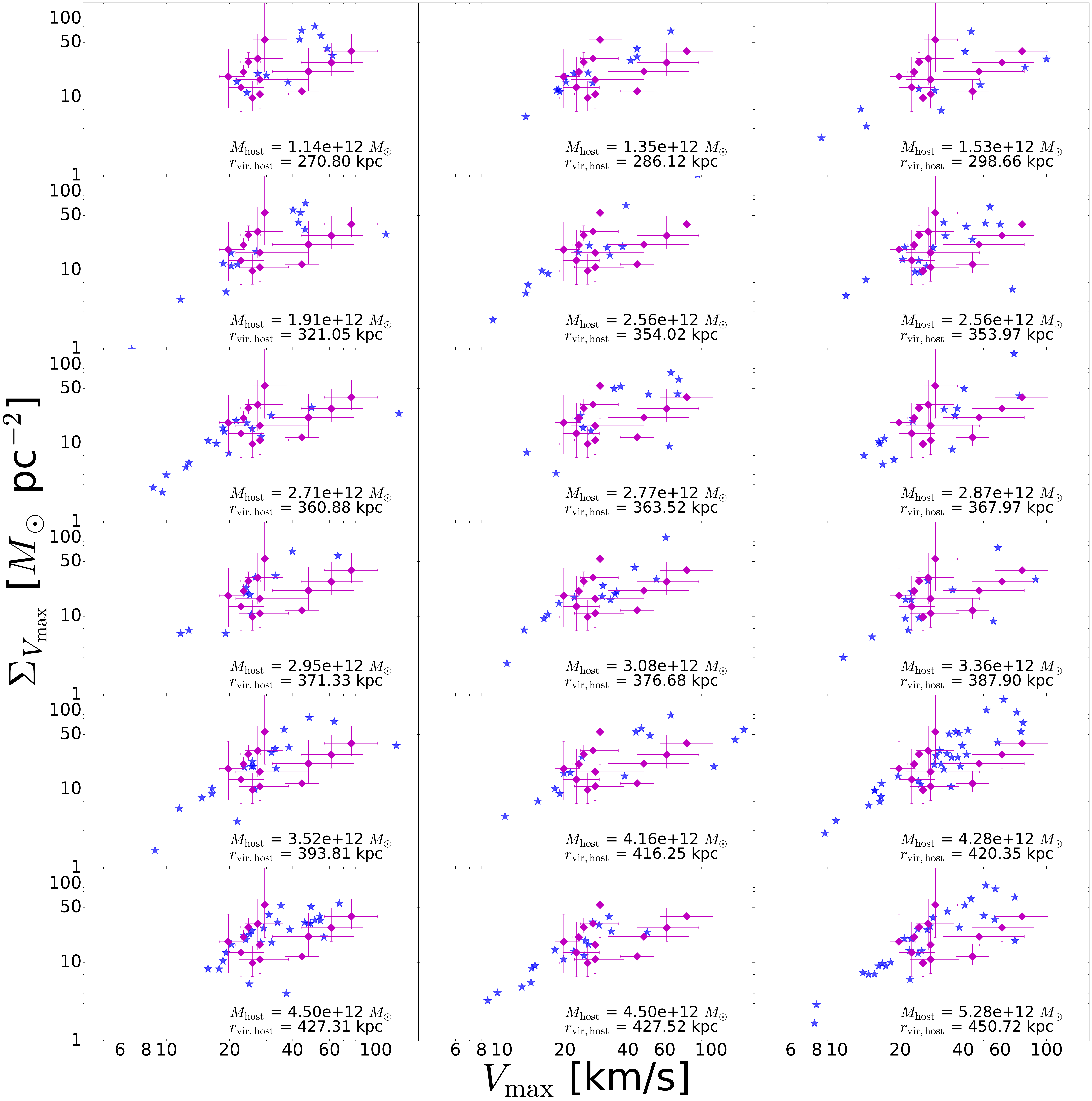}
	\end{center}
    \caption{Same as Figure~\ref{fig1}, but for the 15 selected host halos from the Cosmogrid simulation. The virial masses and radii of host dark halos are indicated in each panel.\label{fig2}}
\end{figure*}

\begin{figure}[t!]
	\begin{center}
	\includegraphics[scale=0.46]{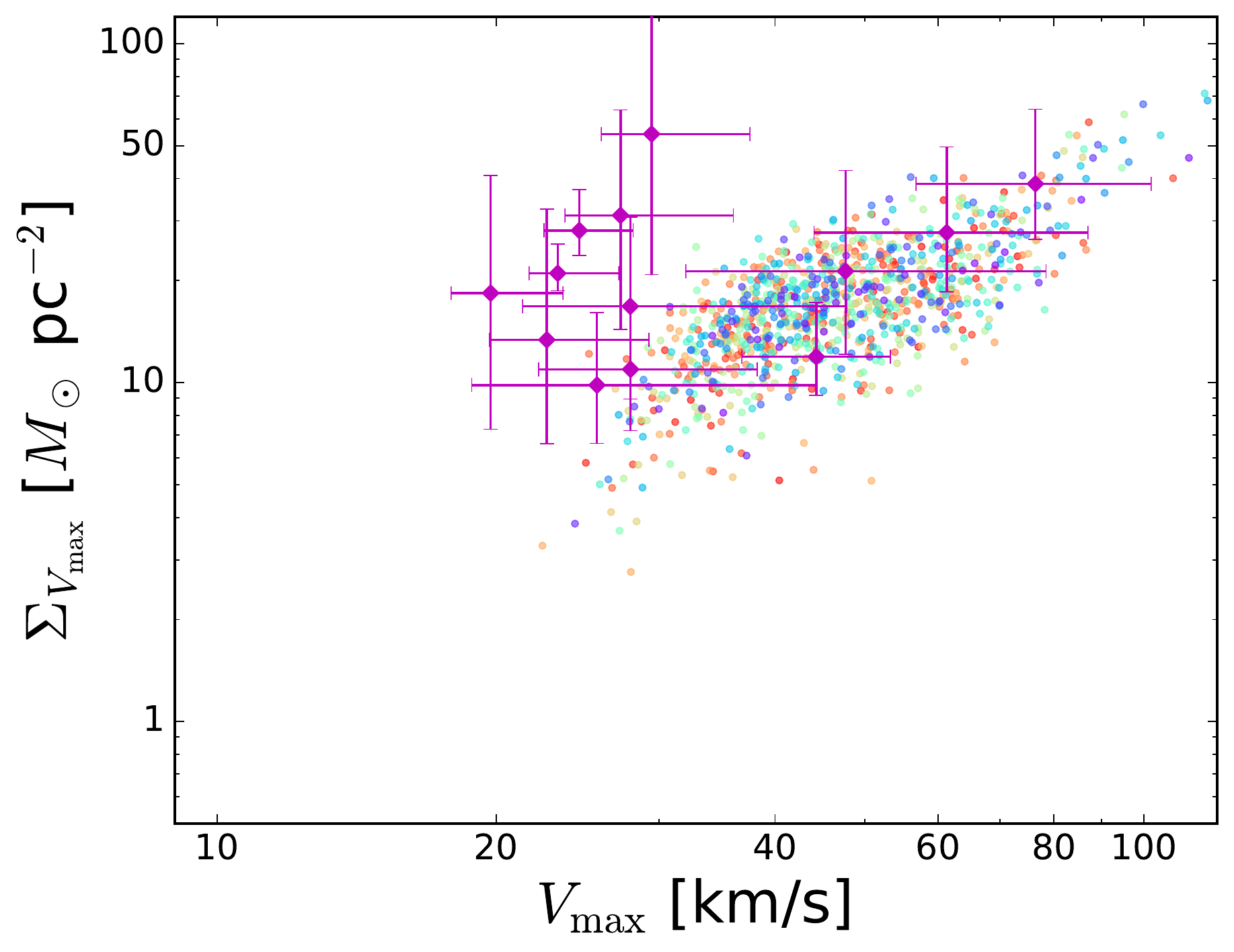}
	\end{center}
    \caption{Same as Figure~\ref{fig1}, but for the Illustris simulations with the highest mass resolution. The different colors of points depict the subhalos associated with the different host halos. \label{fig3}}
\end{figure}

From the above three criteria with respect to the simulated dark halos, we extract the selected subhalos from each host dark halo. 
The number of these subhalos is listed in the fifth column of Table~\ref{tab2}.
In the next section we demonstrate the subhalo distributions on the $\Sigma_{V_{\rm max}}$~versus~$V_{\rm max}$ plane from this dark matter simulation, and compare it with the observed relation.
Moreover, tracking the evolution history of subhalos, we investigate the origin of the observed $\Sigma_{V_{\rm max}}$~versus~$V_{\rm max}$ relation and whether this relation depends on the difference in the properties of each host halo.

\section{The universal dark halo relation for the satellite galaxies}\label{sec:Sigma}

Using $V_{\rm max}$ and $r_{\rm max}$ of subhalos, which fulfill the above three criteria, we calculate their spherical-averaged $\Sigma_{V_{\rm max}}$ derived from equation~(\ref{SVmax}).
First of all, as a indicator representing the compactness of a dark halo, we utilize the mean physical density within the radius of the maximum circular velocity in units of the critical density~($\rho_{\rm crit,0}$) as defined by~\citet[][hereafter D07]{Dieetal2007}:
\begin{equation}
\rho_{V_{\rm max}} = \frac{\overline{\rho}(<r_{\rm max})}{\rho_{\rm crit,0}} = 2\Bigl(\frac{V_{\rm max}}{H_0 r_{\rm max}}\Bigr)^2,
\label{rhoVmax}
\end{equation}
where $H_0$ denotes the Hubble constant at $z=0$.
This corresponds to the physical concentration indicator of a dark halo.
The concentration indicator, which is commonly used in $N$-body simulations, is defined by the ratio between a virial and a scale radii of a halo, assuming an NFW profile.
However, because density profiles of less-massive subhalos, especially those affected by tidal forces from a massive host, should deviate from their initial mass profiles~\citep{AW1986,EHyetal2003,Penetal2008}, the usually defined concentration indicator is not necessarily appropriate.
Thus, we adopt an alternative $\rho_{V_{\rm max}}$ as the concentration indicator of subhalos in this work.

\subsection{Comparison between the observations and the simulations}

Figure~\ref{fig1} and~\ref{fig2} show the results derived from both the simulations and the observations on the $\Sigma_{V_{\rm max}}-V_{\rm max}$ plane focusing on mass scales of dSph-sized dark halos.
It is found from these figures that {\it the dark halo surface densities derived from our simulated dark subhalos agree very well with those from observations, even though we implement only pure dark matter simulations without baryonic effects.}
To confirm this accordance, we also calculate $\Sigma_{V_{\rm max}}$ using the available catalog in the Illustris Project\footnote{http://www.illustris-project.org}, a series of $N$-body and hydrodynamics simulations on cosmological volume. 
We use the results from the highest mass resolution run, 
Illustris-1\footnote{The run parameters of the Illustris-1 simulation are $m_{\rm DM}=6.3\times10^6{\rm M}_{\odot}$ and $\varepsilon_{\rm DM}=1.4$~kpc for a dark matter particle, and $m_{\rm baryon}=1.3\times10^6{\rm M}_{\odot}$ and $\varepsilon_{\rm baryon}=0.7$~kpc for a baryonic particle, respectively. Therefore, this simulation have $\sim830$ ($\sim50$)~times lower dark matter mass resolution than the high-resolution (the Cosmogrid) simulations. The cosmological parameters are adopted by WMAP-9 results~\citep{Hinetal2013}, namely $(\Omega_0,\Omega_{\Lambda},h,n_s,\sigma_8)=(0.2726,0.7274,0.704,0.963,0.809)$.} and select the MW-sized halos and their subhalos though the same criteria above~(i.e., $r_{\rm s}>2\varepsilon_{\rm DM}$ and $M_{\rm vir}\gtrsim10^{9.8}{\rm M}_{\odot}$).  
Figure~\ref{fig3} shows the results of comparison with the Illustris simulation.
From this figure, the effect of mass resolution emerge at low mass scales~($V_{\rm max}\leq30$~km~s$^{-1}$), and it seems like that these $\Sigma_{V_{\rm max}}$ are slightly declined by the baryonic effects at higher mass scales~($V_{\rm max}\geq30$~km~s$^{-1}$).
However, since the dark halo surface densities calculated by the Illustris simulation do not significantly differ from those from observations as with the results of the pure $N$-body simulations, the baryonic feedbacks may have an insignificant effect on this surface density.

This implies that $\Sigma_{V_{\rm max}}$ may be unaffected by net effects of baryon physics and be dependent only on the intrinsic properties of dark halos.
Indeed, a radius of the maximum circular velocity of a dark halo should be much larger than star forming regions, which settle in the center part of a dark halo, and thus this radius would be influenced only by gravitational effects such as mass assembly history and tidal force from their host.
Therefore we suggest that even if inner dark matter profiles at dwarf-galaxy scales can be transformed from cusped to cored due to energy feedback from gas outflows driven by star-formation activity of galaxies, the expelling dark matter particles from the center of a dark halo can not escape beyond the radius of the maximum circular velocity.
In other words, the cusp-to-core transformation, which is a possible solution to the core-cusp and the too-big-to-fail problem, might not affect the $\Sigma_{V_{\rm max}}-V_{\rm max}$ relation.
Interestingly, the recent studies for this transition mechanism can give support to this suggestion~\citep[e.g.,][]{Ogietal2014,OB2015}.
This is, therefore, a possible reason why $\Sigma_{V_{\rm max}}$ of subhalos in numerical simulations accords well with those of observed dSphs, regardless of whether simulations include baryonic physics or not, and also suggests that the universality is an intrinsic property of dark halos.

\subsection{The Universal $\Sigma_{V_{\rm max}}-V_{\rm max}$ relation}

More intriguingly, comparing each panel in Figure~\ref{fig1} and~\ref{fig2}, all host halos have a similar relation between $\Sigma_{V_{\rm max}}$ and $V_{\rm max}$.
To confirm this quantitatively, we perform a least-squares fitting method to the relation, $\alpha$, from $\Sigma_{V_{\rm max}}\propto V_{\rm max}^{\alpha}$ with respect to the subhalos associated with the four host halos in the high-resolution simulations.
As a result, from the gray dashed lines in each panel of in Figure~\ref{fig1}, we obtain $\alpha=0.99$,~$1.13$,~$1.12$, and~$1.14$ for H1, H2, H3, and H4, that is, $\Sigma_{V_{\rm max}}$ is basically proportional to $V_{\rm max}$.
Also, we redo the same fitting procedure to the all subhalos from each simulation and figure out that these subhalos reside along the relation, $\Sigma_{V_{\rm max}} = 0.35V_{\rm max}^{1.11}$ (as seen in the top-left panel in Figure~\ref{fig5}). 
Therefore, it is found that {\it the dark halos associated with a host halo exhibit the remarkable universal relation, irrespective of the difference in each host halo's property such as mass, radius, and concentration}.\footnote{We also investigate whether there is a tight relation between $\rho_{V_{\rm max}}$ and $V_{\rm max}$ and find that $\rho_{V_{\rm max}}-V_{\rm max}$ largely depends on host halo properties and has a much larger scatter than $\Sigma_{V_{\rm max}}-V_{\rm max}$.}
Broadly, $\Sigma_{V_{\rm max}}$ is generally constant across a wide range of $V_{\rm max}$ of $\sim10-400$~km~s$^{-1}$ suggested by HC15a, whilst focusing only on low mass scales with $V_{\rm max}\lesssim100$~km~s$^{-1}$, especially the satellite galaxies, this surface density is actually dependent on $V_{\rm max}$.

\section{What is the origin of the universal $\Sigma_{V_{\rm max}}-V_{\rm max}$ relation?}\label{sec:Evolution}
To understand the origin of the universality of the $\Sigma_{V_{\rm max}}-V_{\rm max}$ relation found for dark halos on the scales of dwarf galaxies, we investigate the evolutionary histories of subhalos in details, especially the link between orbital evolution and $r_{\rm max}$ (and $V_{\rm max}$) evolution, in the Cosmogrid simulations.
The time resolutions of the high-resolution simulations are not enough to investigate the orbital evolutions of their subhalos, thereby we focus only on the subhalos in the Cosmogrid simulations.
Using the merger tree data of subhalos computed by~\ctrees, we trace the evolution of the mass distribution in satellite halos undergoing tidal stripping from a host halo.

\subsection{Evolutionary history of subhalos}
\begin{figure*}[htbp]
  \begin{center}
   \includegraphics[scale=0.6]{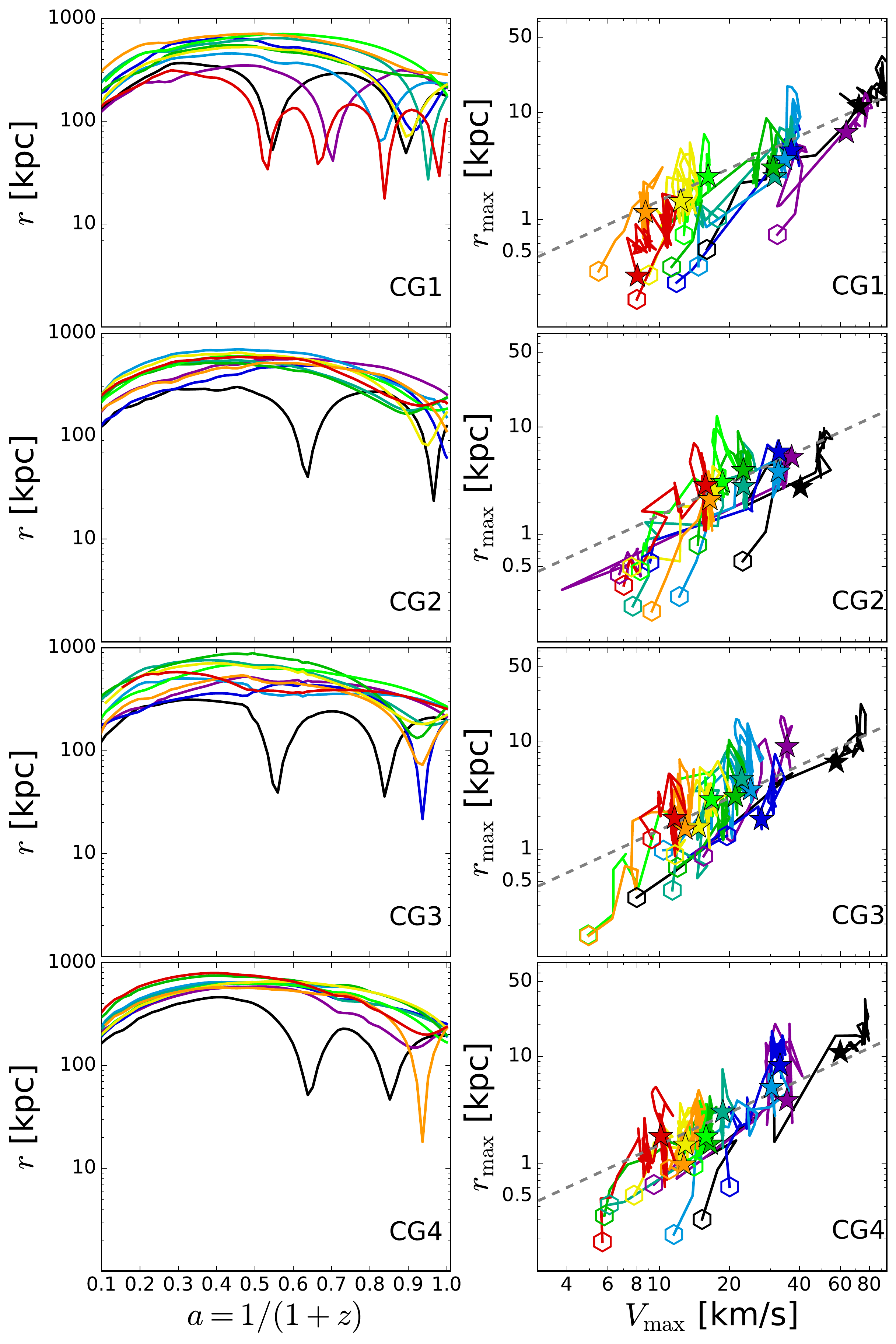}
  \end{center}
 \caption{Evolutionary tracks of subhalos associated with MW-sized dark halo in the Cosmogrid simulation. {\it Left four panels}: Distances to the center of host halo versus scale factor for each host. The different color indicates the ten selected subhalos. {\it Right four panels}: Evolution of subhalo on the $r_{\rm max}-V_{\rm max}$ plane. Each colored line in individual panels is the same subhalos in the corresponding left panels. The open hexagon and the filled stars show starting points~($z=12$) and ending points~($z=0$) of each subhalo evolution, respectively. The dashed lines in each panel indicate $r_{\rm max}\propto V_{\rm max}$ relation.\label{fig4}}
\end{figure*}

The left four panels in Figure~\ref{fig4} show orbital evolutionary tracks of ten representative subhalos associated with four parent dark halos derived from the Cosmogrid simulations.
We choose the four host halos showing relatively rapid growth at a high redshift out of 18 MW-like dark halos, because subhalos associated with these parent halos did fall in at some earlier time and thus might have passed through the pericenter of the host a number of times.
The basic properties of the four Cosmogrid host halos are listed in Table~\ref{tab3}.
In the left panels of Figure~\ref{fig4}, we preferentially select subhalos, which have undergone at least one pericenter passage with respect to a host halo.
Eventually, the majority of subhalos have, however, several or little experiences of pericenter passage.
This implies that the subhalos, which have already passed pericenters many times and thus have been disturbed strongly by tidal effects, have possibly been disrupted or lost their masses significantly by the present day. Such subhalos do not fulfill our subhalo criteria.
Therefore, most of selected subhalos in each panel show the recent infall into their hosts.

The right four panels in Figure~\ref{fig4} display the evolution in the $r_{\rm max}-V_{\rm max}$ plane of subhalos corresponding to those in each left panel.
On this plane, the halo tracks start at the lower-left corner because dark halos are still small at a high redshift. After subhalos move toward upper right during their mass-growth phases, both their $r_{\rm max}$ and $V_{\rm max}$ start decreasing because of their infall into a host halo and associated tidal stripping.
This means that tidal stripping is likely to stop the inside-out dark matter assembly by truncating its outer mass.
It is worth noting from these figures that {\it evolutionary tracks of most subhalos on $r_{\rm max} - V_{\rm max}$ follow virtually the same path, in spite of the differences in host halos and trajectories of subhalos}.
We interpret this interesting evolutionary path as meaning that internal structures of dark halos would be robust against several merger events~\citep[e.g.,][]{Deh2005,Kazetal2006} and the tidal effects on a dark halo might not deviate significantly from $r_{\rm max} - V_{\rm max}$ sequence even though those effects not only truncate its dark matter at the outer region but also affect slightly its inner structures. Thus, dark halos track back their $r_{\rm max}$ and $V_{\rm max}$ to near where ones were before.

Some of subhalos, which experience frequent pericenter passage and/or have small pericenter distances, would deviate somewhat from the universal $r_{\rm max} - V_{\rm max}$ sequence to its upward or downward, because of strong tidal mass loss and tidal disturbance of dark halo structures.
As examples, we focus on the four subhalos in Figure~\ref{fig4}:
(1) the halo, which experiences pericenter passage four times (red line in CG1),
(2) the halo with pericenter passage twice (black line in CG1), and 
(3) the halo, which undergoes infall suddenly and deeply into its host halo (blue line in CG3).
(4) the halo, which undergoes the same as (3)'s halo (orange line in CG4).
Because these subhalos have been truncated tidally, where its outer region is made less bound at every periceter passages, the remnants of such a strong tidal interaction have eventually low densities but with more centrally concentrated.
This is why their $r_{\rm max}$ values are relatively lower than those in the $r_{\rm max} - V_{\rm max}$ sequence and are greatly changed.

\subsection{A possible solution to the origin of the universal relation}
The evolutionay sequence on the $r_{\rm max} - V_{\rm max}$ plane described above and the deviation from it are basically consistent with the results shown in~D07~(see their Fig. 13).
Therefore, it is suggested that subhalos within their own host halos have a common dynamical evolution, regardless of the properties of host halos. 
Moreover, as mentioned in~D07, this sequence in all host halos can be expressed simply as the linear relation, $r_{\rm max}\propto V_{\rm max}$, and we confirm their suggested relation~(dashed lines in the four right panels in Figure~\ref{fig4}).
This relation thus plays a key role in untangling the origin of the universal $\Sigma_{V_{\rm max}}-V_{\rm max}$ relation.
Based on $r_{\rm max}\propto V_{\rm max}$, the mean physical dark halo density defined by equation~(\ref{rhoVmax}) becomes $\rho_{V_{\rm max}} \propto(V_{\rm max}/r_{\rm max})^2 \simeq const$.
On the other hand, $\Sigma_{V_{\rm max}}$ can be rewritten by $\rho_{V_{\rm max}}$ as  
$\Sigma_{V_{\rm max}}\propto\rho_{V_{\rm max}} r_{\rm max}$.
Consequently, we arrive at $\Sigma_{V_{\rm max}}\propto V_{\rm max}$ using the $r_{\rm max}- V_{\rm max}$ relation.

\begin{figure*}[t!]
	\begin{center}
	\includegraphics[scale=0.5]{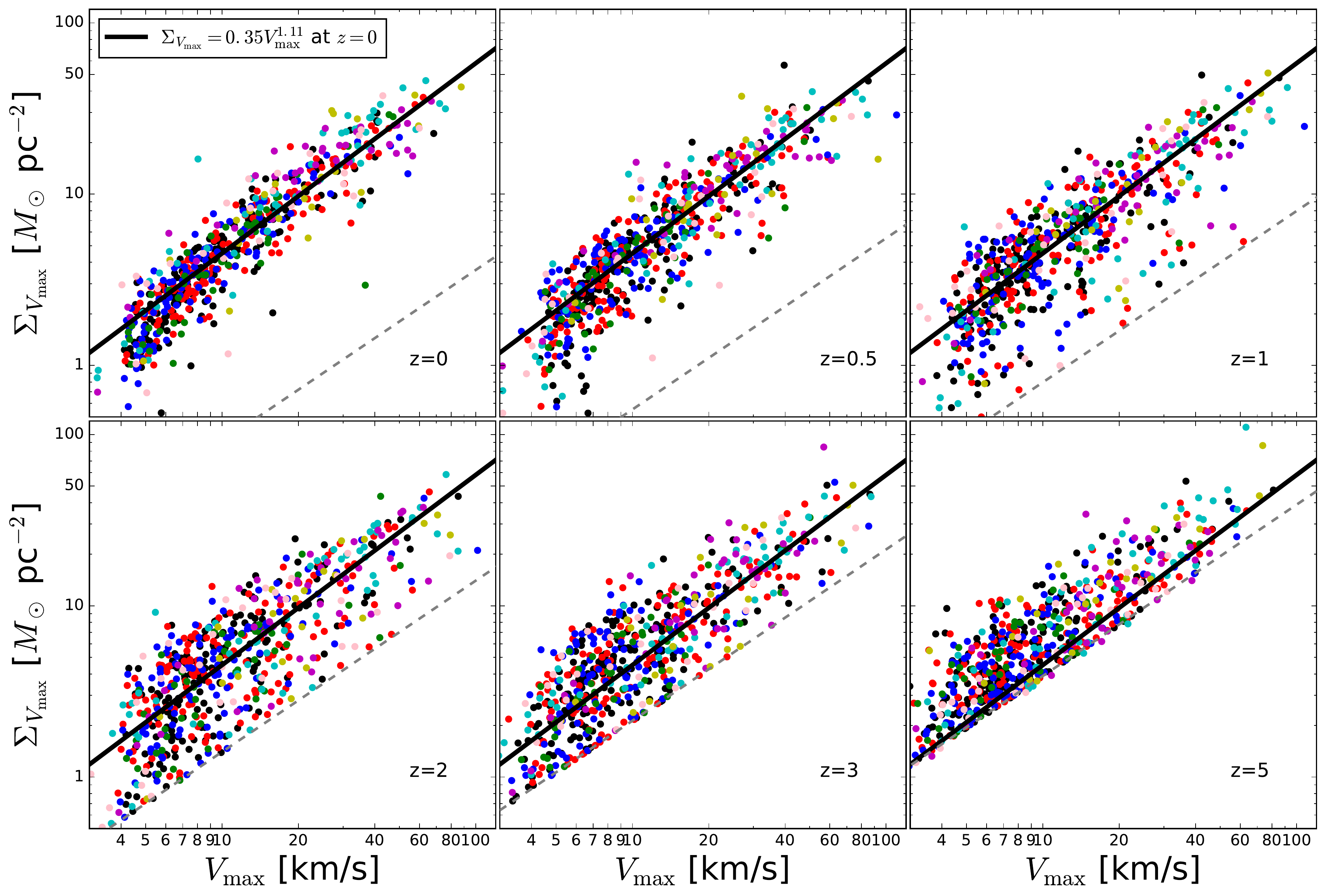}
	\end{center}
    \caption{Redshift evolution of $\Sigma_{V_{\rm max}}-V_{\rm max}$ distribution for the subhalos of all host dark halos that are shown in Table~\ref{tab3}.
     The different colored points depict the subhalos associated with the different host halos. The black solid lines in each panel indicate the fitting result of $\Sigma_{V_{\rm max}}=\beta V^{\alpha}_{\rm max}$ at $z=0$, and the gray dashed lines show the $\Sigma_{V_{\rm max}}-V_{\rm max}$ relations if $r_{\rm max}=r_{\rm vir}$ at each redshift. At higher redshifts, some subhalos would be not stable dynamically, thereby $r_{\rm max}$ cannot be identified clearly and thus are regard as $r_{\rm vir}$. This is why a fraction of subhalos which are along the dashed lines increase with redshifts.\label{fig5}} 
\end{figure*}

Furthermore, subhalos evolve generally along $r_{\rm max}\propto V_{\rm max}$, although the relation holds some dispersion because dark halo structures in these subhalos is affected by a change in mass accretion rate.
Thus, the $\Sigma_{V_{\rm max}}$ versus $V_{\rm max}$ relation, that we have derived here, can be nearly constant with time. 
Following this hypothesis, we plot, in Figure~\ref{fig5}, the redshift evolution of all subhalos associated with the different host halos on the $\Sigma_{V_{\rm max}}-V_{\rm max}$ plane.
Since the virial radius of subhalo decrease naturally with redshifts, the lower limit of $\Sigma_{V_{\rm max}}-V_{\rm max}$ goes up and gets close to $\Sigma_{V_{\rm max}}-V_{\rm max}$ relation at $z=0$.
Although the amplitude of $\Sigma_{V_{\rm max}}-V_{\rm max}$ relation at $z=5$ is thus shifted to upward, its slope does not change significantly.
Therefore, we find that irrespective of some scatters, the relation has indeed been kept roughly invariant since its emergence.

\section{Summary and Conclusions}
We have investigated a dark halo surface density inside a radius at the maximum circular velocity, $\Sigma_{V_{\rm max}}$, first introduced by HC15a, based on the comparison between the results from the observations and those from cosmological $N$-body simulations. 
While the $\Sigma_{V_{\rm max}}$ versus $V_{\rm max}$ relation is semi-analytically inferred from the assumption of an NFW profile combined with an empirical mass concentration relation (HC15a), this can be affected by some environmental effects, such as tidal stripping and heating, as well as halo-to-halo scatters.  
Therefore, in this work, we have scrutinized dark matter halos, especially subhalos associated with a MW-like dark halo, taken from cosmological pure dark matter simulations performed by~\citet{Ishietal2013,Ishietal2016} to understand the properties of this surface density in more details.
We have found the several important results for the properties of the satellite dark halos as follows.
\begin{enumerate}
\item Dark halo surface densities derived from our simulated subhalos associated with MW-sized hosts are in remarkable agreement with those from the observations of the MW and M31 dwarf spheroidal satellites. 
This implies that this surface density is only weakly modified by any baryonic feedbacks.
Therefore, this surface density provides us a clue to understanding fundamental dark halo properties of the Local Group.

\item Even if host dark halos have different masses, compactness, assembly history and the properties of their subhalos, all subhalos are found to obey the universal $\Sigma_{V_{\rm max}}-V_{\rm max}$ relation.
Furthermore, this universality appears to be sustained even at high redshifts.

\item In order to understand the origin of this universal dark halo relation, we have investigated orbital and dynamical evolutions of subhalos.
It is found that most of subhalos have several or little experiences of pericenter passage with respect to a host halo.
Analyzing such a subhalo evolution, we have confirmed that most of subhalos evolve certainly along the $r_{\rm max}\propto V_{\rm max}$ sequence, whereas more disturbed subhalos, which have undergone pericenter passage a number of times and are very close to the center of a host halo, show somewhat deviation from this sequence.
This suggests that tidal force from host halos mainly removes only the outer parts of subhalos, which are accreted from outside, so that this $r_{\rm max}\propto V_{\rm max}$ sequence remains preserved.
Since the $\Sigma_{V_{\rm max}}-V_{\rm max}$ is derived from this sequence, $r_{\rm max}\propto V_{\rm max}$, it is found to be a basic property to understand the common dark halo surface density scales. 
\end{enumerate}

Following the results obtained here, there is the possibility that this universal relation for dark halos exists not only for the MW and M31 but also larger gravitational systems such as galaxy groups and clusters.
For instance, recently discovered ultra diffuse galaxies in clusters~\citep[e.g.,][]{vanetal2015a,vanetal2015b,Kodetal2015,Yagetal2016}, might be a system affected by strong tidal disturbances, and if so, these galaxies may show some characteristic trends in the $\Sigma_{V_{\rm max}}$ versus $V_{\rm max}$ relation.
Also, it remains yet unclear whether the ultra-faint galaxies in the MW indeed obey this universal relation, because of the paucity of observable bright stars in such faint systems.
We thus require high-quality photometric and spectroscopic data of these galaxies.
The current and future facilities such as the Hyper Suprime Cam~\citep{Miyetal2012} and the Prime Focus Spectrograph~\citep{Tametal2016} to be mounted on the Subaru Telescope and high-precision spectroscopy the Thirty Meter Telescope~\citep{Simetal2016} will have the capability to obtain severer constraints on dark halo structures of these galaxies, and enable us to understand the basic properties of dark matter in the universe.

\acknowledgments
We would like to give special thanks to Matthew G. Walker for giving us the kinematic data of Ursa Minor dwarf galaxies and for useful discussions.
This work acknowledges support from MEXT Grant-in-Aid for Scientific Research on Innovative Areas, No.~16H01090 (for K.H.), No.~15H01030 (for T.I.), No.~15H05889, No.~16H01086 (for M.C.).
This research was supported by MEXT as ``Priority Issue on Post-K computer'' (Elucidation of the Fundamental Laws and Evolution of the Universe) and JICFuS. 
GO acknowledges funding from the European Research Council (ERC) under the European Union's Horizon 2020 research and innovation programme (grant agreement No. 679145, project 'COSMO-SIMS').
Finally, Kavli IPMU is supported by World Premier International Research Center Initiative (WPI), MEXT, Japan.



\end{document}